\documentstyle [prl,aps,twocolumn,psfig]{revtex}
\newskip\humongous \humongous=0pt plus 1000pt minus 1000pt
\def\caja{\mathsurround=0pt}
\def\eqalign#1{\,\vcenter{\openup1\jot \caja
	\ialign{\strut \hfil$\displaystyle{##}$&$
	\displaystyle{{}##}$\hfil\crcr#1\crcr}}\,}
\newif\ifdtup

\relax

\begin{document}

\newcommand{\newc}{\newcommand}

\newc{\be}{\begin{equation}}
\newc{\ee}{\end{equation}}
\newc{\ba}{\begin{eqnarray}}
\newc{\ea}{\end{eqnarray}}
\newc{\bea}{\begin{eqnarray}}
\newc{\eea}{\end{eqnarray}}
\newc{\D}{\partial}
\newc{\ie}{{\it i.e.} }
\newc{\eg}{{\it e.g.} }
\newc{\etc}{{\it etc.} }
\newc{\etal}{{\it et al.}} 

\newc{\ra}{\rightarrow}
\newc{\lra}{\leftrightarrow}
\newc{\no}{Nielsen-Olesen }
\newc{\tp}{'t Hooft-Polyakov }
\newc{\lsim}{\buildrel{<}\over{\sim}}
\newc{\gsim}{\buildrel{>}\over{\sim}}
\title{Core phase structure of cosmic strings and monopoles}

\bigskip

\author{M. Axenides}
 
\address{Institute of Nuclear Physics, N.C.S.R. Demokritos,
153 10, Athens; Greece\\
e-mail: axenides@gr3801.nrcps.ariadne-t.gr}

\author{L. Perivolaropoulos and T.N. Tomaras}
  
\address{Department of Physics and Institute of Plasma Physics,
University of Crete and FO.R.T.H. \\ 
P.O.Box 2208, 710 03 Heraklion, Crete; Greece\\
e-mail: leandros@physics.uch.gr, tomaras@physics.uch.gr}

\date{\today}
\maketitle

\begin{abstract}
Global and local symmetries may or may not be restored inside 
topological defects depending upon the values of the parameters of
the model. A detailed study of this parameter dependence of the
core structure of strings and monopoles is presented in the
context of simple models.   

\end{abstract}

\narrowtext


\section{Introduction}
\noindent

A lot of theoretical and experimental effort has been
devoted over the past twenty years to the search for 
defects predicted in field theory models 
and to the study of their dynamics and role 
in particle physics and cosmology \cite{coleman75} \cite{vs94}.
Despite of the fact that todate there is no direct evidence of any 
such object, their potential 
importance in cosmology (structure formation,
baryogenesis \cite{vs94})
and the interest in the non-perturbative structure of field theories 
has provided enough motivation for theorists to search for all possible
extended solutions and to analyze
their physical properties.
The absolutely stable topological
solitons
were studied first \cite{topological}, 
and were followed by the discovery of the metastable 
{\it semilocal} defects\cite{va91}, \cite{preskill92}, 
the {\it electroweak strings} \cite{nambu}, 
and the {\it ribbons} \cite{bt} in
realistic models of high energy physics, including topologically 
trivial ones.
 
The role of solitons is known to depend to some extent on the detailed
profile and the unbroken symmetries in their interior. 
For instance, the cosmological
evolution of a superconducting string \cite{witten} 
differs significantly
from that of a normal string \cite{v81}. 
The purpose of this note is
to study the parameter dependence of the core structure of strings
and monopoles.
As a first step, we restrict ourselves to models which are
simple enough to be analyzed in detail, and which 
furthermore could easily be
embedded into larger realistic theories.
This analysis is particularly interesting in the context of Cosmology,
where due to the temperature dependence of the parameters of
field theories, one may
encounter during the cosmological evolution "phase transitions" 
inside the cores of topological defects.
In fact, laboratory experiments on $^3He$,
designed to investigate the physics of phase transitions
in the Early Universe, have explicitly provided us with 
strong experimental evidence for defect-core transitions 
in the interior of vortices
which appear in the superfluid $^3He-B$ phase \cite{gv87}.

In section II we study the core structure of cosmic strings. The analysis is
done in the context of a U(1) gauge model with 
two classically relevant parameters and with
the original semilocal string as a limiting case. 
A similar analysis is performed in section III where the
't Hooft-Polyakov monopole is embedded in an SU(2) gauge model with an
extended Higgs sector. The parameter space is divided in regions
corresponding to the two possibilities of the monopole-core global symmetry.  
Possible extensions and applications are commented upon 
in the final discussion section.

\section{Embedded Nielsen-Olesen Vortex}

Consider the simple
extension of the Abelian Higgs model with two equally 
charged scalars $\Phi_1$ and $\Phi_2$
described by the lagrangian density  
\begin{equation}
{\cal {L}}=-{1\over 4} B_{\mu \nu} B^{\mu \nu} + 
{1 \over 2} (D_\mu \Phi)^\dagger D^\mu \Phi - V(\Phi_1, \Phi_2)
\label{model1}
\end{equation}
where $\Phi =(\Phi_1, \Phi_2)$, $B_{\mu\nu}=\partial_\mu B_\nu-
\partial_\nu B_\mu$, 
$D_\mu = \partial_\mu - i g B_\mu$, and with $\tau_3$ the third Pauli
matrix
\begin{equation} 
V(\Phi_1, \Phi_2)=-{M^2 \over 2}\Phi^\dagger \Phi - 
{m^2 \over 2} {\Phi^\dagger}\tau_3 \Phi + 
{\lambda \over 4} ({\Phi^\dagger}\Phi)^2
\label{potential1}
\end{equation}
For generic values of the parameters the vacuum has $\Phi_1^* \Phi_1 = 
(M^2+m^2)/\lambda$ and $\Phi_2 = 0$. The model is an example
of a U(1) gauge theory with the gauge group broken spontaneously 
to the identity, and as such it supports the existence of  
topologically stable strings. $\Phi_1$, the field with non-vanishing 
vacuum expectation value, carries their winding and vanishes at the
center. As for $\Phi_2$, it vanishes at infinity, but its profile 
inside the string has no geometric or topological constraints and 
is determined dynamically by the field equations.
In this section we will study in detail the structure of the 
string core and how it varies
with the parameters of the model. 
One might imagine (\ref{model1}), (\ref{potential1}) 
embedded in a more realistic theory
with $\Phi_2$ coupled
to electromagnetism. A non-zero core value of 
$\Phi_2$ in this case would render the string 
superconducting with well known observable effects \cite{witten}. 

Of the four parameters of the model, one sets the scale, another 
may be pulled outside of the action to play the role of the semiclassical
parameter $\hbar$ and
there remain two classically relevant ones. Specifically, 
by rescaling fields and 
coordinates according to
\begin{equation}
\Phi_{1(2)} \to M \Phi_{1(2)} /\sqrt{\lambda}, 
\;\; B_\mu \to M B_\mu/g, \;\; x_i \to x_i/M
\end{equation} 
one is left with the parameters
\begin{equation}
\alpha \equiv {m \over M} \;\;{\rm and}\;\; 
\beta \equiv {g\over \sqrt{\lambda}}.
\end{equation}
Various limiting cases of (\ref{model1}), (\ref{potential1}) have 
been studied before.
First, for both $\alpha$ and $\beta$ equal to zero a simple scaling argument 
shows that the model 
does not possess any kind of stable defect solution.
Second, for $\alpha=0$ one obtains an SU(2)$_{global} \times$
U(1)$_{local}$ symmetric model,
with the U(1) gauge field distinguishing an S$^1$ fiber out of the S$^3$
vacuum. The "strength" of this fibration is proportional to $\beta$. 
For $\beta > \beta_0 \equiv \sqrt{2}$
the embedded Nielsen-Olesen configuration i.e. with $\Phi_2=0$ 
is a classically stable solution of the model \cite{h92}, and is the first
example of a semilocal vortex studied before \cite{va91}.   
For smaller values of $\beta$ an instability arises due to the development
of a non-vanishing $\Phi_2$ inside the core and the string 
blows up to the vacuum.
Finally, the global model with $\beta=0$ was studied recently in \cite{ap97}.
In analogy to the U(1) gauging of the previous limiting case, 
one may think of the term $|\Phi_1|^2-|\Phi_2|^2$ in (\ref{potential1}) as
defining an S$^1$ "scalar" fibration with "strength" $\alpha$ 
on the S$^3$ vacuum manifold of the $\alpha=0$ model. 
Topologically stable global strings (with logarithmically infinite 
energy per unit length) exist in this case 
for all positive values of $\alpha$. Furthermore, for $\alpha > 
\alpha_0 \simeq 0.4$ they
have $\Phi_2=0$, while $\Phi_2 \neq 0$ when $\alpha < \alpha_0$. 
 
The simplest guess for the parameter dependence of the 
core structure of the strings in (\ref{model1}), (\ref{potential1}) 
is then the following:
In the $(\alpha, \beta)$ plane there is a curve connecting the critical
points $(\alpha_0, 0)$ and $(0, \beta_0)$ on the two axes, 
outside of which
the embedded \no strings are stable, and inside of which the stable strings
are characterized by non-vanishing $\Phi_2$. This picture will indeed be
confirmed below and the curve $\alpha_{crit}(\beta)$ will
be determined numerically.

The ansatz for the string solutions with winding number
$n$ has the axially symmetric form
\begin{equation}
\Phi_1 = f(\rho) e^{i n \theta}\, , \;\; \Phi_2 = G(\rho)\, , \;\;
{\vec B} = {\hat e}_\theta \;{B(\rho)\over{\rho}}
\label{string-ansatz}
\end{equation}
and the corresponding energy functional is in units of $M^4/{2 \lambda}$
\begin{equation}
\eqalign{
&E=\int d^2x\Bigl[f'^2 + G'^2 + {1\over \beta^2} {{B'^2}\over {\rho^2}} 
+ {{(n-B)^2} \over {\rho^2}} f^2 \cr 
&+ {{B^2}\over {\rho^2}} G^2
+{1\over 2} (f^2 + G^2)^2 - (1+\alpha^2)f^2 - (1-\alpha^2) G^2 \Bigr]
\label{string-energy}
}
\end{equation}
It leads to the following field equations:
\begin{equation}
\eqalign{
f'' + {f' \over \rho} - {(n-B)^2 \over \rho^2} f +
(1+\alpha^2-f^2 -G^2) f &= 0 \cr
B'' - {B' \over \rho} + \beta^2 f^2 (n-B)  - \beta^2 G^2 B &= 0 \cr
G'' + {G' \over \rho} -{B^2 \over \rho^2} G + (1-\alpha^2 - f^2 -G^2)G &= 0
\label{string-equations}
}
\end{equation}
We restrict our analysis to the minimal $n=1$ case. In the
end we comment briefly on the results for higher values of $n$. 
Finiteness of the energy and the field equations at the origin
imply the boundary conditions
\begin{equation}
\eqalign{
f(\infty) &= \sqrt{\alpha^2 +1},
\;\;\; G(\infty) = 0, 
\;\;\; B(\infty) = n \cr
f(0)&= 0, \;\;\; G^\prime (0)=0,
\;\;\; B(0) = 0
\label{string-bc}
} 
\end{equation}
For all values of the parameters $\alpha$ and $\beta$, equations 
(\ref{string-equations}) admit the embedded \no solution with $G(\rho)=0$.
As explained above, this is known to be classically 
stable on the $\alpha$-axis for
$\alpha > \alpha_0$ and on the $\beta$-axis for $\beta >\beta_0$.
For arbitrary ($\alpha$, $\beta$), the region of stability of
the corresponding solution is determined by the requirement that
in the expansion of the energy functional around it the first non-trivial
term in $\delta E$ is strictly positive for all field variations.
To quadratic order in field changes $\delta E = \delta E_{NO} + \delta E_G$,
where $\delta E_{NO}$ is independent of $G(\rho)$ and identical in form 
with the perturbation obtained for the topologically stable Nielsen-Olesen
vortex. Thus $\delta E_{NO} > 0$ for all $\delta f$ and $\delta A$
and may be ignored.
$\delta E_{G}$ on the other hand represents the quadratic correction
to the energy of the Nielsen-Olesen string due to the excitation of $G(\rho)$.
It may be written in the form:
\be
\delta E_{G}=\int d^2x \;G \,{\hat O}\, G
\ee
with
\be
{\hat O} = -{1 \over \rho} {d \over {d\rho}} \rho {d \over {d\rho}} + 
{{B^2} \over
\rho^2} + f^2 + \alpha^2 - 1
\ee
where $f$, $B$ are the \no fields obtained by solving the first two equations
of system (\ref{string-equations}) with $G=0$.
For stability of the embedded \no vortex we require that $\delta E_G > 0$
for arbitrary small perturbation $G$.
This is equivalent to demanding that the operator ${\hat O}$ has no negative
eigenvalues. 
The region of stability of the embedded \no vortex was determined 
numerically.
The method is straightforward. We pick a point in the ($\alpha$, $\beta$)
plane, set $G=0$ and solve the remaining equations (\ref{string-equations})
to find the corresponding \no solution. We then construct the
operator ${\hat O}$ and consider the eigenvalue problem
\be
{\hat O}\, G = \omega^2 \,G
\label{stability1}
\ee
\begin{figure} 
\centerline{
\psfig{figure=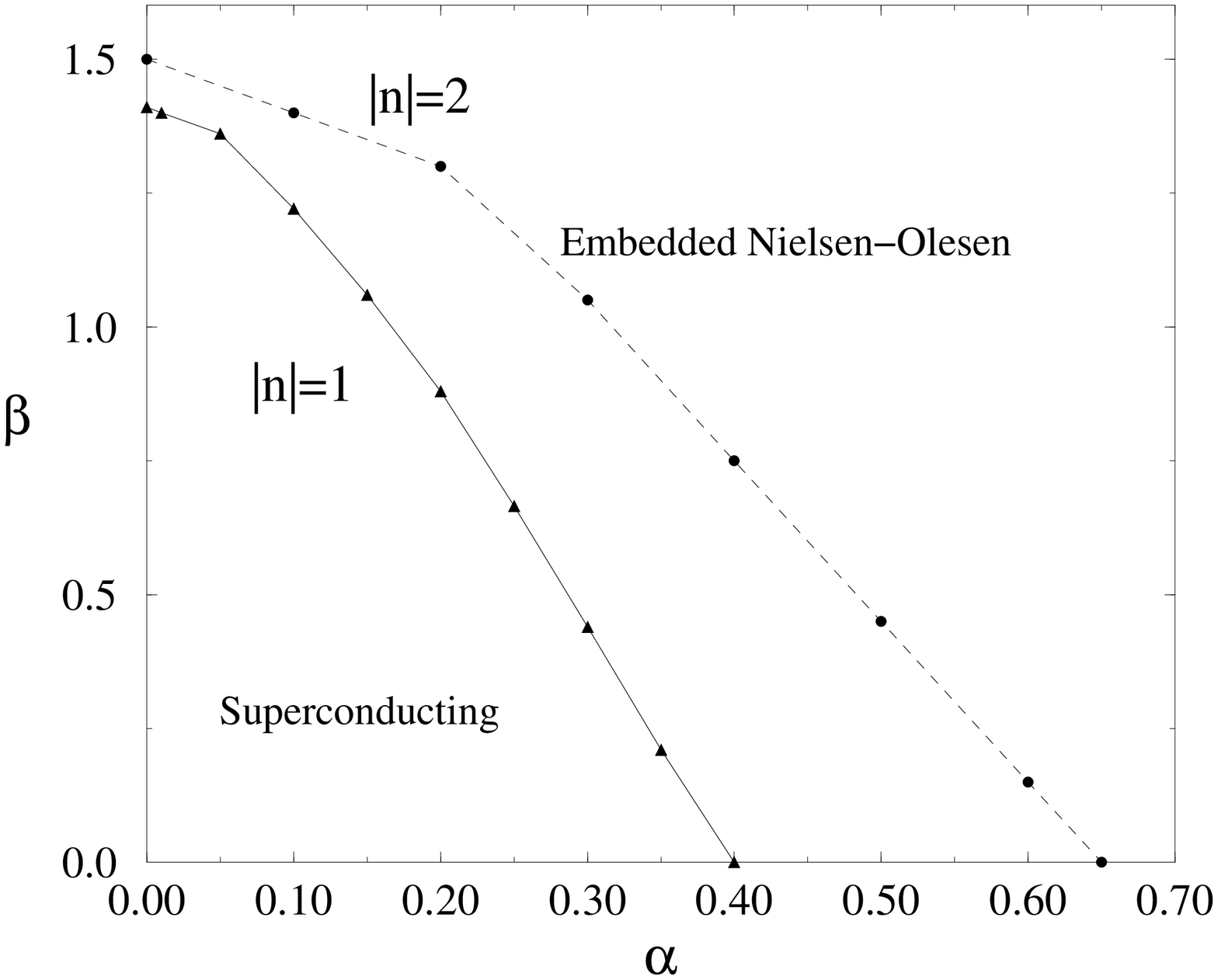,height=3.2in,angle=-90}
\psdraft
}
Figure 1:  The string-core structure for all values of parameters
$\alpha, \beta$. Inside the region bounded by the solid line $\Phi_2 \neq 0$,
while outside it $\Phi_2$ vanishes.
\end{figure}
We use a fourth order Runge-Kutta routine
with initial
conditions $G(0) = 1$, $G'(0)=0$ to investigate if there is a bound 
state in equation (\ref{stability1}). 
If there is a  $\rho_0$ such that for $\rho >
\rho_0$ we have $G(\rho) < 0$ with the above initial condition, then
there is clearly a bound state for the considered values of $\alpha,
\beta$. This implies an instability of the embedded \no
vortex towards another stable configuration with non-zero order
parameter in the core. 
In Figure 1 we plot the parameter space $(\alpha, \beta)$ and
display the regions of stability and instability of the
embedded \no vortex. 
We have repeated the above analysis to the $|n|=2$ case. As shown in Figure 1
increasing $|n|$ leads to an expansion of the superconducting region.
This is related to the fact that the behaviour of $f$ near zero is
$f \sim r^{|n|}$. Thus increasing $|n|$ leads to a broadening of the
potential of the Schroedinger operator ${\hat O}$ and favours the existence
of negative eigenvalues.    

Figures 2 and 3 contain the results of the numerical integration
via a relaxation method of the system (\ref{string-equations}).
In Figure 2 we show a stable embedded \no configuration 
and in Figure 3 an example of a vortex solution for 
parameter values in the superconducting region. The
latter has a non-zero order parameter in the core.
For $\alpha \neq 0$ this relaxed configuration is stable due to a
non-trivial topology. For $\alpha = 0$ there is no topological stability
and the \no configuration spreads its flux to infinity \cite{h92}. 
\begin{figure}
\centerline{
\psfig{figure=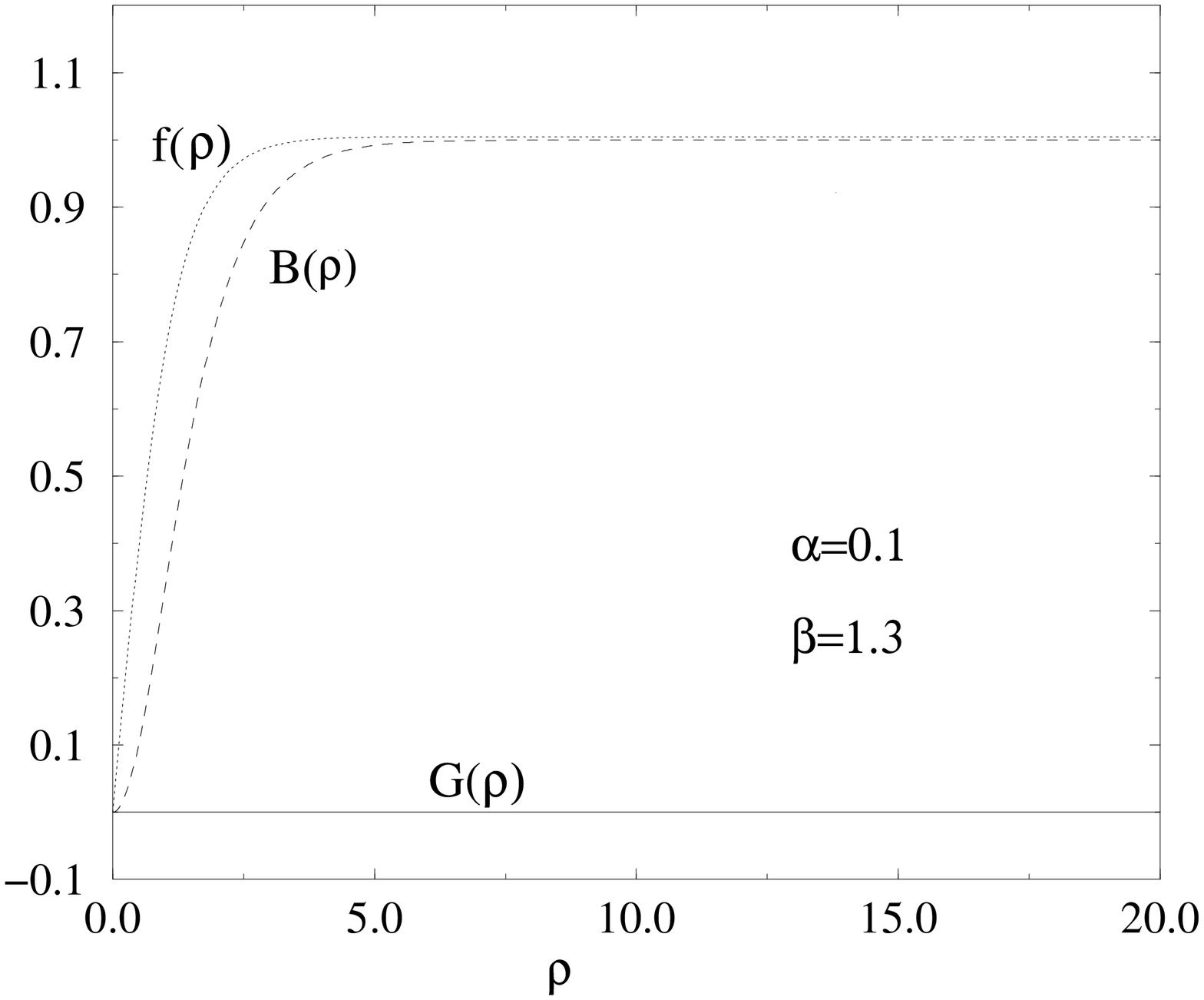,height=3.2in,angle=-90}
\psdraft
}
Figure 2: For $\alpha = 0.1$ and $\beta = 1.3$ the stable string 
configuration shown here is the embedded \no solution.
\end{figure}
\begin{figure}
\centerline{
\psfig{figure=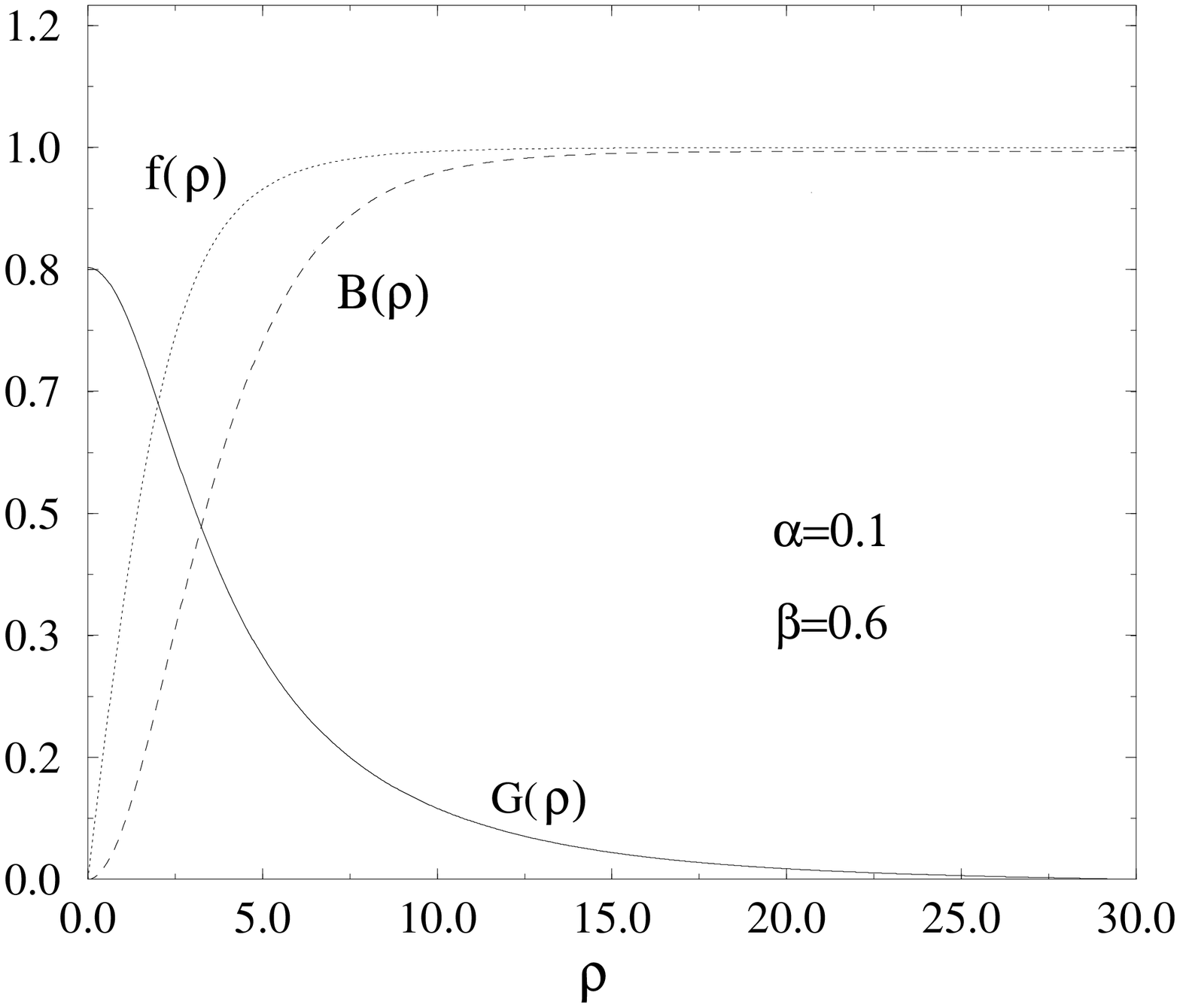,height=3.2in,angle=-90}
\psdraft
}
Figure 3: The profile of the stable string for 
$\alpha = 0.1$ and $\beta = 0.6$. It has $\Phi_2 \neq 0$ inside the core.
\end{figure}

To leading order in the temperature $T$ the parameter $\alpha$ becomes
$T$-dependent $\alpha^2(T)=-m^2 / [-M^2+ (2 \lambda + g^2) T^2/4]$, 
while $\beta$
is constant. Strings form at $T$ slightly below 
the critical temperature $T_C$. $\alpha^2(T \sim T_C)$ is very large
and the strings start off normal. 
When T is such that $\alpha(T) = \alpha_{crit}(\beta)$ for the given
$\beta$ the string becomes superconducting.

\section{Stability of the Embedded Monopole}
It is straightforward to apply the above method to the study
of the fate of global symmetries inside the monopole 
core. We choose to work in the context of the simple model
\begin{equation}
{\cal L} = -{1\over 4} F^a_{\mu \nu}F^{a\mu \nu} + 
{1\over 2} D_\mu \Phi^a D^\mu \Phi^a + {1\over 2} (\D_\mu \Phi^4)^2 - V
\label{model2}
\end{equation}
describing the dynamics of an O(3) gauge field $A_{\mu}^a$ 
coupled to the scalar triplet $\Phi^a$ $a$=1,2,3, which in addition 
interacts with the gauge singlet $\Phi^4$ through the potential
\be
V = {\lambda\over 4} (\Phi^a \Phi^a + \Phi^4 \Phi^4 - v^2)^2 - {m^2
\over 2}(\Phi^a \Phi^a - \Phi^4 \Phi^4 )
\ee
The field strength and the covariant derivative are given by 
$F^a_{\mu \nu} = \D_\mu A ^a_\nu - \D_\nu A^a_\mu +
g \epsilon^{abc}A^b_\mu A^c_\nu$ and $D_\mu \Phi^a \equiv \D_\mu \Phi^a + 
g \epsilon^{abc} A^b_\mu \Phi^c$, respectively.  

The above is a simple extension of the Georgi-Glashow O(3) 
model with the two classically relevant parameters 
\be
\tilde\alpha={m \over {\sqrt{\lambda} v}} \;\; {\rm and} \;\; 
\tilde\beta = {g\over \sqrt{\lambda}}
\ee
as revealed after the rescalling 
$x^i \to x^i /(v \sqrt{\lambda})$, 
$\Phi^a \to v \Phi^a$, $\Phi^4 \to v \Phi^4$, and 
$A^a_\mu \to v A^a_\mu$.

For $\tilde\alpha=\tilde\beta=0$ (\ref{model2}) possesses an
O(4) global symmetry. A non-zero $m$ 
breaks this symmetry explicitly to the gauged O(3) subgroup,
and this in turn is spontaneously broken to O(2) 
by the vacuum of the model. 
According to the
standard lore the model admits for generic values of 
$\tilde\alpha$ and $\tilde\beta$ a whole tower of
topologically stable magnetic monopoles. 
The minimal one in its "ground state" has the
spherically symmetric form 
\begin{equation}
\eqalign{
\Phi^a &= \delta_{i a} {x^i \over r} f(r), \;\; 
A^a_i = \epsilon_{aij} {x^j \over r} W(r) \cr   
\Phi^4 &= G(r) 
\label{monopole-ansatz} 
}
\end{equation}
with $f(r)$ and $W(r)$ necessarily vanishing at the origin $r=0$.
Depending on the profile of $G(r)$, which as in the string case 
can only be decided by solving the field equations, 
the symmetry inside the monopole core will be either O(3) 
($G(r)\neq 0$) or the full O(4) ($G(r)=0$). Which one of the two 
is realized depends on the values of the parameters of the model.
Using the same numerical method as in the string case we
give below a complete map of the parameter space based on the core
symmetry of the corresponding magnetic monopole. 

It is convenient to define $K(r) \equiv 1-\tilde\beta r W(r)$, in which case
the field equations for the three unknown functions of the ansatz
take the form:
\begin{equation}
\eqalign{
f'' + {{2f'}\over r} - {{2f}\over r^2} K^2 +
(1+\tilde\alpha^2-f^2 - G^2) f &= 0 \cr
K^{\prime \prime} - {{K(K^2 -1)}\over{r^2}} - \tilde\beta^2 f^2 K &= 0 \cr
G'' + {{2G'}\over r} + (1-\tilde\alpha^2-f^2 - G^2) G &= 0 
\label{monopole-equations}
} 
\end{equation}
while the corresponding boundary conditions, dictated by the finiteness
of the energy and the field equations at the origin, are 
\begin{equation}
\eqalign{
f(\infty) &= \sqrt{1+\tilde\alpha^2}, \;\;\;
G(\infty) = 0, \;\;\; K(\infty) = 0 \cr
f(0) &= 0, \;\;\;\;\;\;\;\;\;\;\;\;\;\;\;\; G^\prime (0) = 0, 
\;\;\;\;\; K(0) = 1
\label{bc2}
}
\end{equation}
\begin{figure}
\centerline{
\psfig{figure=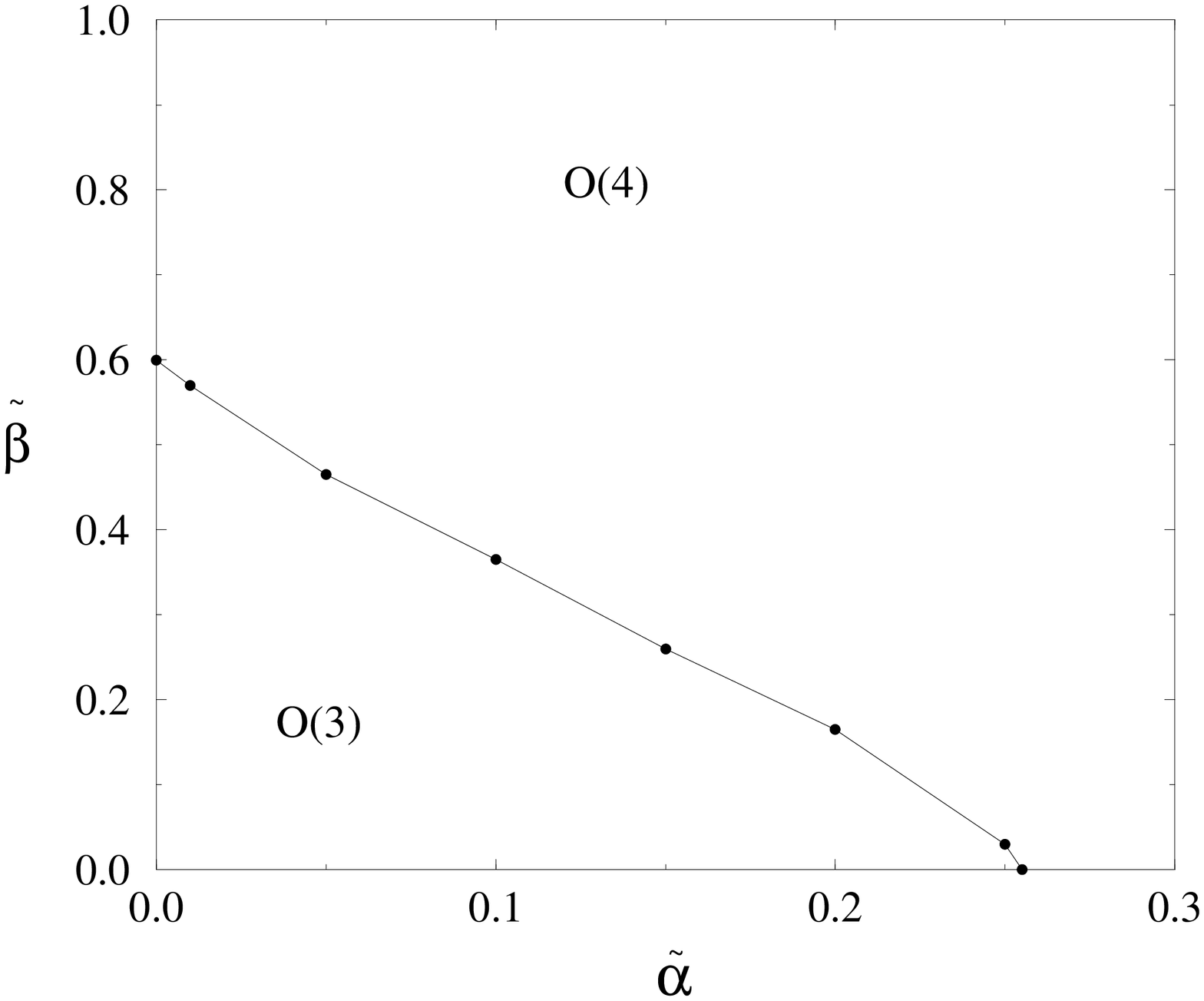,height=3.2in,angle=-90}
\psdraft
}
Figure 4: The  dependence on the parameters ${\tilde \alpha}$ and ${\tilde \beta}$ of the monopole-core global
symmetry. 
\end{figure}
We now look for instability modes of the embedded 
\tp monopole solution (i.e. with $G(r)=0$
at all $r$) of equations (\ref{monopole-equations}). 
Using the same approach as in the case of the \no vortex we obtain the
linearized eigenvalue problem corresponding to the last equation in 
(\ref{monopole-equations})
\be
-G'' - {{2 G'} \over r} + (f^2 - 1 + \tilde\alpha^2) G = \omega^2 G
\label{stability2}
\ee
where $f$ is obtained from the \tp monopole solution i.e. solving the
system of the first two equations in (\ref{monopole-equations}) with
$G=0$. Notice that contrary to the string case there is no coupling of the
gauge field to $G$. In models where such coupling exists and the
gauge symmetry is $O(4)$ the embedded \tp monopole is always
unstable due to the Brandt-Neri-Coleman
instability \cite{bn79}. This instability is realized as a screening of
the long range magnetic field of the monopole by the other 
massless gauge fields of the theory \cite{bvb94}. No such instability 
exists in the model under discussion because the gauge symmetry is
O(3) and not O(4). 
\begin{figure}
\centerline{
\psfig{figure=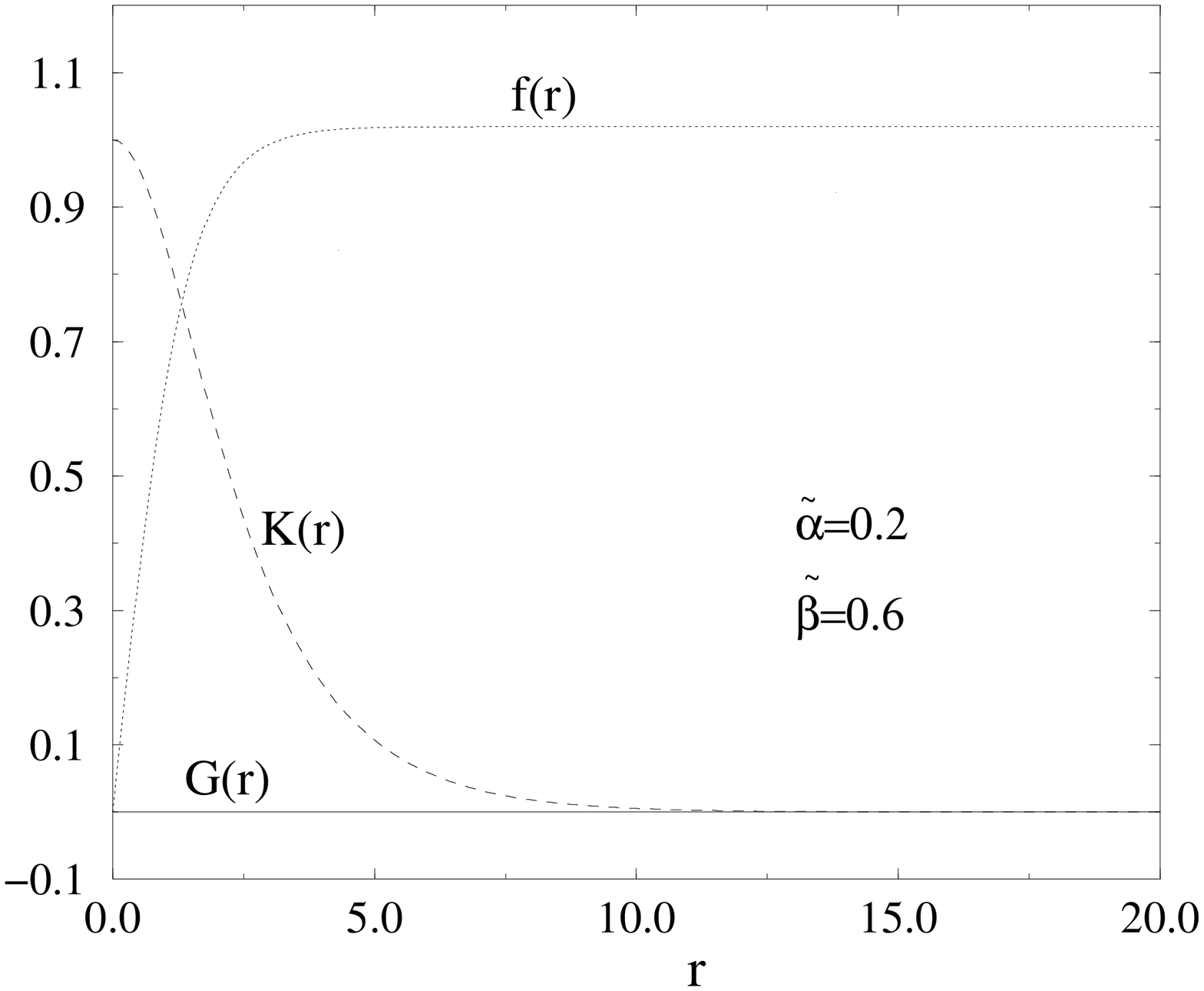,height=3.2in,angle=-90}
\psdraft
}
Figure 5: For $\tilde\alpha = 0.2$ and $\tilde\beta = 0.6$ 
the stable monopole is the embedded 't Hooft-Polyakov 
solution with G(r)=0.
\end{figure}
We have solved the eigenvalue problem (\ref{stability2}) 
for several parameter pairs
($\tilde\alpha, \tilde\beta$). The parameter space shown in Figure 4 
is divided into two regions according to the symmetry properties of
the core of the stable monopole solution.
For strong scalar and gauge fibrations i.e. for
$\tilde\alpha, \tilde\beta$ large, the core symmetry of the stable
monopole solution is O(4).
Negative modes
(instabilities) towards an O(3) symmetric core develop for 
weak fibrations as in the case of the
embedded \no vortex.
The embedded \tp monopole in the semilocal limit was first 
discussed in \cite{preskill92}. Our results
confirm the qualitative analysis presented there. 

We solved numerically equations (16) and (17) using a relaxation method
for a variety of parameter values.
In Figure 5 an example of a stable embedded \tp monopole solution is
shown, while in
Figure 6 we plot the profile of a stable monopole solution
with O(3) symmetric core.
  
\begin{figure}
\centerline{
\psfig{figure=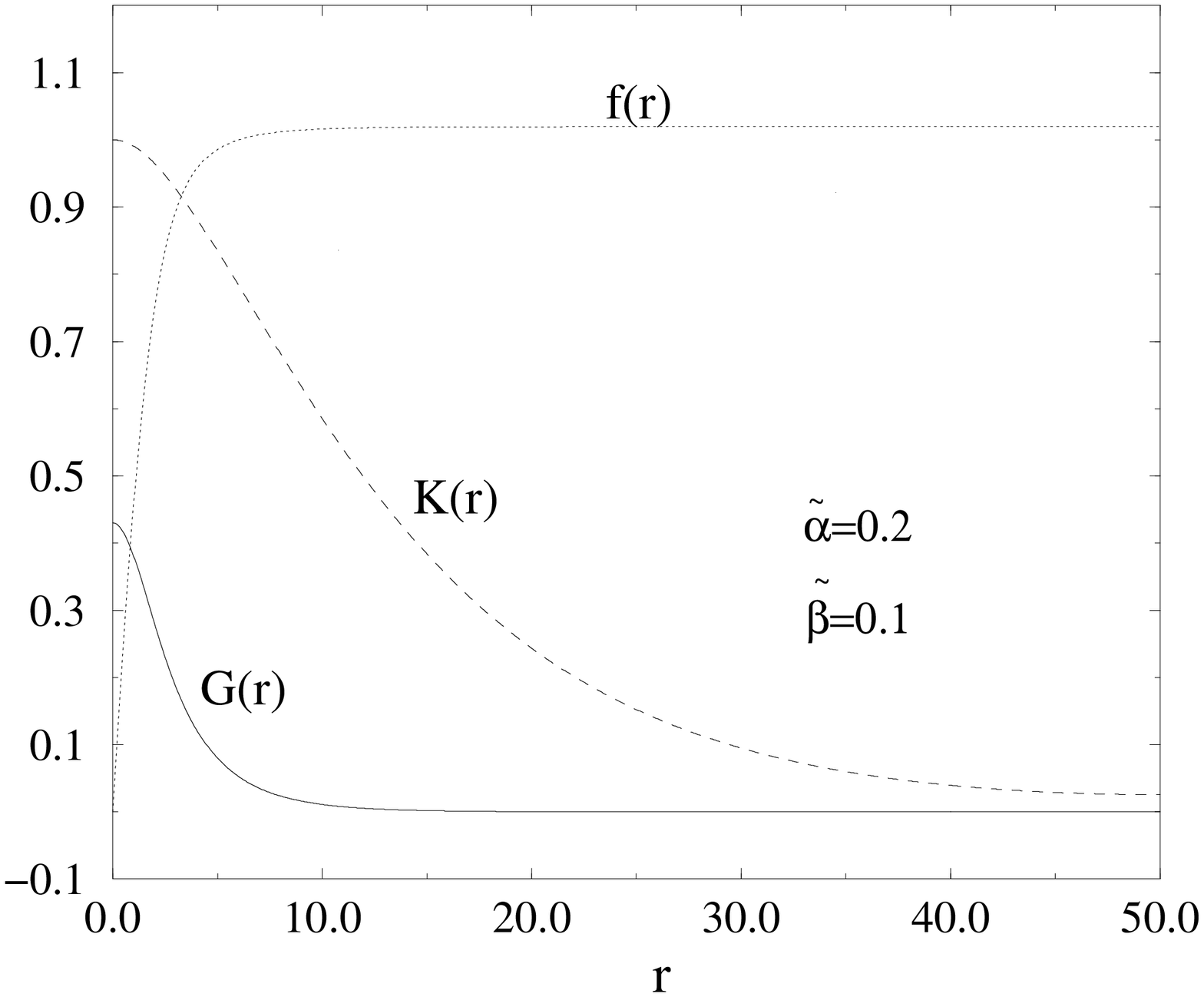,height=3.2in,angle=-90}
\psdraft
}
Figure 6: The stable monopole shown here for $\tilde\alpha = 0.2$
and $\tilde\beta = 0.1$ has $\Phi^4$ excited inside the core.
\end{figure}

\section{Conclusion}

The dependence of the core phase structure of flux vortices 
and magnetic monopoles was studied explicitly
in the context of simple gauge models carrying these
topological defects. The combination of "scalar" and
"gauge" fibrations on the vacuum manifold leads to
instabilities of the core of the embedded \no vortices and
\tp monopoles, which result in a non-trivial interior
in the stable solution. 
The models discussed here are very simple and allow for only
two phases in the corresponding defect interiors. 
A larger variety of core phases is of course expected in realistic
models with more Higgs multiplets and richer symmetry pattern.
The cosmological implications of the transitions between 
these different core possibilities deserve further study.

\section{Acknowledgements}
We would like to thank Drs. A. Achucarro and T. Vachaspati for discussions.

This work is the result of a network supported by the European Science
Foundation. 
The European Science Foundation acts as catalyst for the development of 
science by bringing together leading scientists and funding agencies to
debate, plan and implement pan-European initiatives.
This work was also supported by the EU grant CHRX-CT94-0621 as well 
as by the Greek General Secretariat of Research and Technology grant
$\Pi$ENE$\Delta$95-1759.

\vfill
                      
\eject

\end{document}